%% file: pitaunu.tex
\documentclass[ aps, twocolumn, superscriptaddress, showpacs, preprintnumbers, amsmath, amssymb ]{revtex4-1}
\pdfoutput=1

\usepackage{graphicx} 
\usepackage{dcolumn}  
\usepackage{subcaption}

\usepackage{esvect}

\usepackage{hyperref}

\newcommand{\bzero}{\ensuremath{B^0}}
\newcommand{\pitaunu}{\ensuremath{\bzero \to \pi^- \tau^+ \nu_{\tau}}}
\newcommand{\EE}{\ensuremath{e^+e^-}}
\newcommand{\BB}{\ensuremath{B\bar{B}}}
\newcommand{\qq}{\ensuremath{q\bar{q}}}
\newcommand{\unit}[1]{\ensuremath{~\mathrm{#1}}}
\newcommand{\cunit}[2]{\ensuremath{~\mathrm{#1}#2}}
\newcommand{\br}[1]{\ensuremath{\mathcal{B}\left(#1\right)}}
\newcommand{\btag}{\ensuremath{B_{\rm tag}}}
\newcommand{\bsig}{\ensuremath{B_{\rm sig}}}
\newcommand{\bvis}{\ensuremath{B_{\rm vis}}}
\newcommand{\mbc}{\ensuremath{M_{\rm bc}}}
\newcommand{\pbeam}{\ensuremath{p_{\rm beam}}}
\newcommand{\pmiss}{\ensuremath{p_{\rm miss}}}
\newcommand{\mmiss}{\ensuremath{M_{\rm miss}}}
\newcommand{\absdr}{\ensuremath{|dr|}}
\newcommand{\absdz}{\ensuremath{|dz|}}
\newcommand{\eecl}{\ensuremath{E_{\rm ECL}}}
\newcommand{\pvalue}{$p$-value}
\newcommand{\lncntnbout}{\ensuremath{\ln({o_\mathrm{tag}^\mathrm{cs}})}}
\newcommand{\cntnbout}{\ensuremath{o_\mathrm{tag}^\mathrm{cs}}}

\newcommand{\vecsq}[1]{\ensuremath{\vv{p}^{2}}}

\hypersetup{%
	pdfauthor={Philipp Hamer},
	pdftitle={Search for B0 -> pi tau nu with hadronic tagging at Belle}
}

\begin{document}
\title{Search for {\boldmath $\pitaunu$} with hadronic tagging at Belle}

\input{author}

\begin{abstract}
We search for the process $\pitaunu$ using the full Belle data set of $711\,{\rm fb}^{-1}$,
corresponding to $772 \times 10^6 B\bar{B}$ pairs,
collected at the $\Upsilon(4S)$ resonance with the Belle detector at the KEKB asymmetric-energy $e^+ e^-$ collider.
We reconstruct one $B$ meson in a hadronic decay and search for the $\pitaunu$ process in the remainder of the event.
No significant signal is observed and an upper limit of $\br{\pitaunu} < 2.5 \times 10^{-4}$ is obtained at the $90\%$ confidence level.
\end{abstract}


\maketitle

{\renewcommand{\thefootnote}{\fnsymbol{footnote}}}
\setcounter{footnote}{0}

\section{\label{sec:intro}Introduction}
The decay $\pitaunu$~\footnote{Throughout this paper, the inclusion of the charge-conjugate mode process is implied unless otherwise stated.}
is mediated by the $W^+$ boson via the $\bar{b} \to \bar{u}$ transition.
The transition amplitude is described by~\cite{Khodjamirian_pilnu_2011}
\begin{align}
  \left< \pi^{-}|u\gamma_{\mu}\bar{b}|B^0\right> &= f^+(q^2)\left[ 2 p_\mu + \left( \frac{1 - m^2_{B} - m^2_{\pi}}{q^2} \right) q_\mu \right] \nonumber \\
																	&+ f^0(q^2) \frac{m^2_{B} - m^2_{\pi}}{q^2} q_\mu,
\label{eq:scalardecayamplitude}
\end{align}
with $p$ and $q$ being the momentum transfers to the pion and lepton pair, respectively.

The form factors $f^+$ and $f^0$ can be computed from QCD light-cone sum rules \cite{Duplancic_btopi_2008,Khodjamirian_pilnu_2011} for $q^2 < 16\cunit{GeV^2}{/c^4}$
and lattice QCD \cite{Dalgic_lattice,*Gulez_dalgic_lattice_erratum,Bailey_lattice,Lattice:2015tia} for $q^2 > 16\cunit{GeV^2}{/c^4}$.
Various parametrizations exist to interpolate between the two regions.
In this study, we use the parametrization introduced by Bourrely, Caprini, and Lellouch (BCL)~\cite{bcl}, which can describe both form factors in $m_\tau^2 \leq q^2 \leq \left(m_B - m_\pi\right)^2$.
The parameter values are taken from Ref.~\cite{Khodjamirian_pilnu_2011}.

It has been stated \cite{Khodjamirian_beauty_pos,Khodjamirian_pilnu_2011} that the differential ratio
\begin{equation}
  \frac{\mathrm{d}\Gamma \left(B \to \pi \tau \nu_{\tau}\right)/ \mathrm{d}q^2}{\mathrm{d}\Gamma \left(B \to \pi \ell \nu_{\ell}\right) / \mathrm{d}q^2},\ \ \ \ell = e,\;\mu
  \label{eq:ratio_tau_ell}
\end{equation}
can be used as a test for the Standard Model (SM) as it depends solely on the ratio of the scalar and vector form factors $f^0 / f^+$.
The CKM matrix~\cite{Kobayashi01021973} element $|V_{ub}|$ enters both differential branching fractions and cancels in the ratio.

In new physics models like the two-Higgs-doublet model (2HDM)~\cite{Lee:1973iz,Branco:2011iw}, the decay \pitaunu{} can also be mediated by a charged Higgs boson.
Possible contribution of a $H^+$ and other couplings in the 2HDM and MSSM~\cite{Wess:1974tw,Golfand:1971iw}, which would affect the branching fraction and the differential ratio of branching fractions, have been evaluated in Refs.~\cite{Khodjamirian_pilnu_2011} and \cite{dutta_pitaunu,Grossman:1994ax,Chen:2006nua,Kim:2007uq}.

The decay \pitaunu{} has not been observed, nor has an upper limit on the branching fraction been obtained. 
Recent results~\cite{Lattice:2015tia} on the two form factors obtained from a joint fit to (2+1)-flavor lattice QCD calculations and $B \to \pi \ell \nu$ data from Belle~\cite{Ha:2010rf,alexei_bulnu} and BaBar~\cite{delAmoSanchez:2010af,Lees:2012vv} result in 
$\br{ \pitaunu } / \br{ B^0 \to \pi^- \ell^+ \nu_{\ell}} = 0.641(17)$ and 
$\br{ \pitaunu } = 9.35(38) \times 10^{-5}$~\cite{daping_water_pitaunu}.

The signal decay is reconstructed in the four one-prong decays of the $\tau$ lepton, $\tau^- \to \ell^- \bar{\nu}_{\ell} \nu_{\tau}$ with $\ell = e$ or $\mu$, $\tau^- \to \pi^- \nu_{\tau}$, and $\tau^- \to \rho^- \nu_{\tau}$, corresponding to $72\%$ of all $\tau$ decays~\cite{PDG-2014}.
The most powerful decay modes are the two aforementioned hadronic $\tau$ decays and the $\tau^- \to e^- \bar{\nu}_{e} \nu_{\tau}$ mode, while the $\tau^- \to \mu^- \bar{\nu}_{\mu} \nu_{\tau}$ decay mode does not improve the final expected significance.
This is mainly due to low muon momenta in the signal decay and the resulting low muon identification efficiency.
The result of this analysis is based on the three most powerful $\tau$ decay modes.

\section{\label{sec:datasample}Data Sample}
The search for \pitaunu{} described in this paper is performed on the full data sample collected with the Belle detector
at the KEKB asymmetric-energy $\EE$ (3.5 on 8.0~GeV) collider~\cite{KEKB,*[][ and references therein]Abe:2013kxa}, operating at the $\Upsilon(4S)$ resonance.
The data sample consists of an integrated luminosity of $711\unit{fb^{-1}}$, which corresponds to $(771.6 \pm 10.6) \times 10^6 \BB$ pairs.

The Belle detector is a large-solid-angle magnetic
spectrometer that consists of a silicon vertex detector (SVD),
a 50-layer central drift chamber (CDC), an array of
aerogel threshold Cherenkov counters (ACC),  
a barrel-like arrangement of time-of-flight (TOF)
scintillation counters, and an electromagnetic calorimeter (ECL)
comprised of CsI(Tl) crystals located inside
a superconducting solenoid coil that provides a 1.5~T
magnetic field.  An iron flux-return yoke located outside of
the coil is instrumented to detect $K_L^0$ mesons and to identify
muons (KLM).
Two inner detector configurations were used. A $2.0\unit{cm}$ beampipe
and a 3-layer SVD were used for the first sample
of $152 \times 10^6 B\bar{B}$ pairs, while a $1.5\unit{cm}$ beampipe, a 4-layer
silicon detector and a small-cell inner drift chamber were used to record
the remaining $620 \times 10^6 B\bar{B}$ pairs~\cite{svd2_natkaniec,*svd2_ushiroda}.
The detector is described in detail in Ref.~\cite{Belle,*[ also see detector section in ]Brodzicka:2012jm}.

The study was performed as a blind analysis based on simulated data.
Monte Carlo (MC) samples were generated with EvtGen~\cite{EvtGen}
and the detector simulation was performed by GEANT3~\cite{gsim}.
Recorded beam background was added to the MC samples.
The expected non-beam background is estimated using MC samples that describe all physics processes at Belle.
A resonant $\Upsilon(4S)$ event at Belle produces a \BB{} pair.
Two samples of $b \to c$ decays for $\bzero \bar{\bzero}$ and $B^+ B^-$ events, respectively, each contain 10 times the integrated luminosity of the data sample.
Semileptonic $b \to u$ decays are simulated in a sample containing 20 times the integrated luminosity.
Rare $b \to s$ and other rare decays are described in another sample corresponding to 50 times the integrated luminosity of the data.
Continuum $\EE \to \qq$ $(q = u,d,s,c)$ was generated with PYTHIA~\cite{pythia} and included in the analysis in an MC sample containing five times the integrated luminosity of the data sample.
Additionally, a high statistics sample of $\bzero \to X_u \tau \nu$ containing $24 \times 10^6$ events was generated with a phase-space and ISGW2 \cite{isgw2} model.

The signal MC sample is generated using BCL results for the vector and scalar form factors \cite{Khodjamirian_pilnu_2011}.
A total of $84 \times 10^{6}$ $\bzero \bar{\bzero}$ events were generated
with one meson decaying into the signal final state and the other decaying generically.

No constraints on the $\tau$ decay were applied.
The signal MC sample corresponds to approximately 2000 times the expected $\br{\pitaunu} = 9.35 \times 10^{-5}$.

\section{Event Selection}
The complete reconstruction of the $B$ meson decay into the signal final state (\bsig) is not possible due to the presence of at least two neutrinos.
However, since the initial state of the \EE{} collision is completely defined by the momenta of the colliding leptons, we can constrain the signal side by fully reconstructing the other $B$ meson (\btag) in hadronic decay modes.
Tracks and clusters in the event that are not assigned to the \btag{} after the successful reconstruction are assumed to originate from \bsig.

\subsection{Tag side}
This analysis uses the Belle hadronic full-reconstruction algorithm~\cite{ekpfullrecon} based on the artificial Neural Network package NeuroBayes~\cite{neurobayes}.
Neural networks were trained to reconstruct $B^0$ and $B^+$ candidates from a total of 1104 decay channels.

Additional event shape variables are added to suppress continuum events.
$B$ mesons of resonant events are nearly at rest in the center-of-mass (c.m.) frame, leading to a spherical distribution of their decay products.
Continuum events, on the other hand, produce back-to-back jets due the large available kinetic energy.
Useful observables that differentiate between the two event types are the thrust axis of the \btag{} meson~\cite{Brandt:1964sa} and modified Fox-Wolfram moments~\cite{fw,*[the modified version used in this paper is described in ]sfw}.
For this analysis, the thrust axis and the second modified Fox-Wolfram moment are included in the neural network for the full hadronic reconstruction.
If the algorithm does not succeed in reconstructing a $B^0$ candidate, the event is discarded.

Differences in the full reconstruction efficiency between MC and data, depending on the network output and \btag{} reconstruction channel, are observed~\cite{alexei_bulnu}
but depend on the tag-side reconstruction only; a correction factor is determined from charmed semileptonic decays of the signal-side $B$ meson.

\begin{figure*}[htb]
	\includegraphics[width=.8\textwidth]{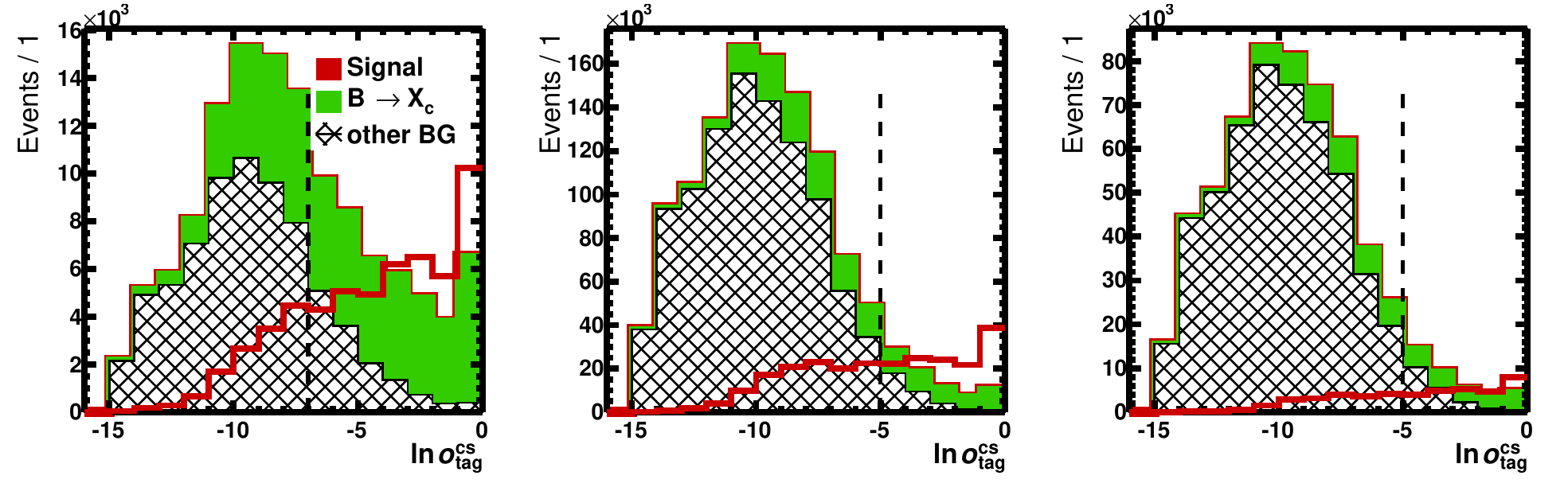}
	\\
	\includegraphics[width=.8\textwidth]{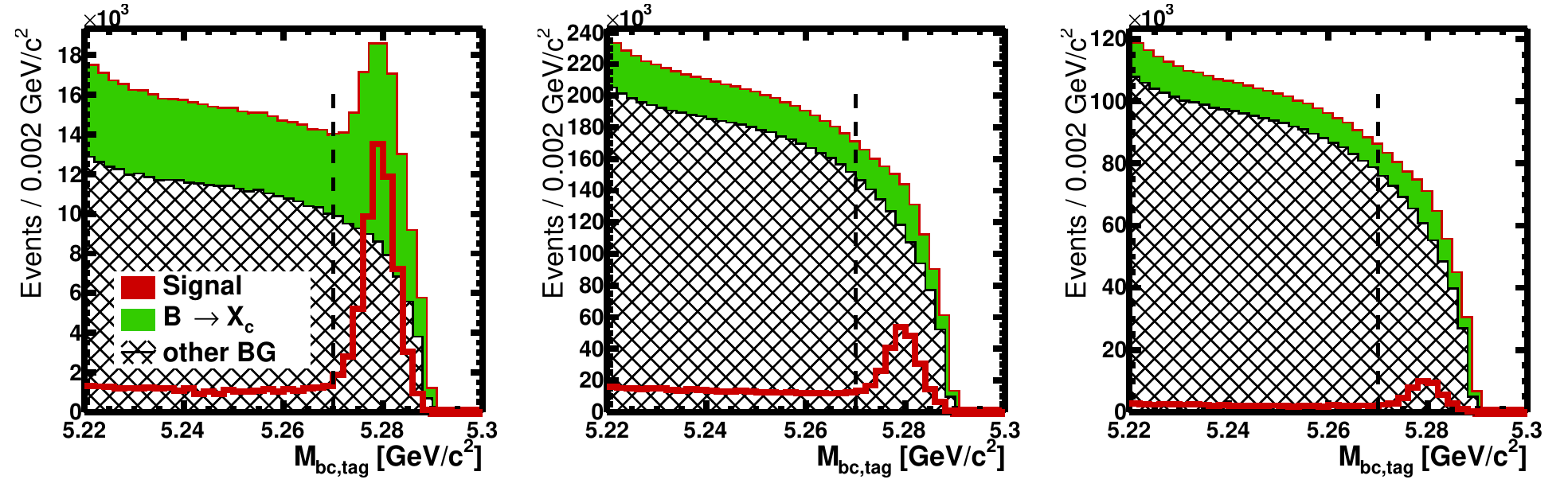}
 \caption{Distributions of \lncntnbout (top) and \mbc{} (bottom) of the tag side candidate for the three reconstruction channels $\tau \to e$ (left), $\tau \to \pi$ (middle) and $\tau \to \rho$ (right) for signal and background, produced from Belle simulation.
  A requirement of $\mbc > 5.27\cunit{GeV}{/c^2}$ is applied in the left plot but no further requirements are applied other than successful reconstruction into one of the three modes.
	The histograms are produced from MC samples, normalized to the data sample size. 
	The signal histogram is scaled by an arbitrary factor of $2000$ for better visibility.
	The dashed vertical line indicates the minimal value required in the final selection; see Section~\ref{sec:evtsel:final}.}
\label{fig:tagside}
\end{figure*}

The beam-energy-constrained mass,
\[
  \mbc = \sqrt{ E^{2}_{\rm beam} - (\vec{p}_{\btag} c)^2 }\; / c^2,
\]
is required to be greater than $5.27\cunit{GeV}{/c^2}$, where $E_{\rm beam}$ and $\vec{p}_{\btag}$ denote the beam energy and reconstructed three-momentum, respectively, of the \btag{}, evaluated in the \EE{} c.m. frame.
With this requirement and the correction factor applied, we estimate a reconstruction efficiency of $0.18\%$ from the signal MC sample, which is in very good agreement with the reconstruction efficiency of $B^0$ mesons in the Belle data sample~\cite{ekpfullrecon}.
The neural network output, $\cntnbout{} \in [0,1]$, is a continuous variable whose high (low) values correspond to candidates which are likely (unlikely) to be a true $B$ meson.
It is used at a later selection stage, as described below.
The distributions of \mbc{} and \lncntnbout{} for the three reconstruction channels are shown in Fig.~\ref{fig:tagside}, where the green (solid) contribution shows the dominant charmed $B \to X_c$ background decays.
No further requirements are applied to the \mbc{} distributions, while $\mbc > 5.27\cunit{GeV}{/c^2}$ is required for the \lncntnbout{} distributions.

\subsection{Signal side}
Only one-prong decays of the $\tau$ lepton are considered in this search.
For of a correctly reconstructed \btag{}, there should be exactly two remaining oppositely-charged tracks in the detector.
Additionally, the event should contain undetected (missing) momentum.
Since the initial state of the \EE{} collisions is given by the four-momenta of the colliding leptons, the undetected momentum can be measured.
The missing momentum is defined as
\[
  \pmiss = 2 \pbeam - p_{\btag} - p_{\bvis},
\]
where $2 \pbeam = p_{e^+} + p_{e^-}$ is twice the beam momentum and \bvis{} denotes the visible part of the \bsig{} meson.
Tracks with low transverse momentum $p_t$ can curl in the solenoidal field and be detected as two tracks with opposite charge.
Any two tracks with $p_t < 275\cunit{MeV}{/c}$ with an angle between the momentum vectors, calculated at the points closest to the interaction point, below $15^\circ$ and total momentum difference less than $100\cunit{MeV}{/c}$ are therefore counted as a single track.
We reduce the number of poor quality tracks by requiring that $\absdr < 2.0\unit{cm}$ and $\absdz < 4.0\unit{cm}$, where \absdz{} and \absdr{} are the distances of closest approach of a track to the interaction point along the $z$-axis and in the transverse plane, respectively.

%
Electron identification \cite{eid} is performed by calculating a likelihood ratio using the matching of charged tracks with the shower position in the ECL, the shower shape, the ratio of the energy deposited in the ECL and the measured momentum, the energy loss $dE/dx$ in the CDC, and the Cherenkov light production in the ACC\@.
Muon identification \cite{muid} is also done by evaluating a likelihood ratio.
Clusters in the KLM are matched to charged tracks by extrapolation.
For matched tracks, the difference between expected and measured penetration depth and the transverse deviation of all KLM hits associated with the track are used in this likelihood ratio.
For a charged track not identified as a lepton, a kaon veto is applied using a likelihood ratio that discriminates between kaons and pions \cite{pid}.
The ratio is formed from the energy loss $dE/dx$ in the CDC, flight time information from the TOF, and photon yield in the ACC.
All remaining tracks are identified as a pion.
Neutral pions are reconstructed from pairs of photons.
The absolute difference between the invariant mass of the $\pi^0$ candidate and the nominal $\pi^0$ mass, normalized to its uncertainty, must be below $3.0$.
Photons are required to have energies in the laboratory frame greater than $50\unit{MeV}$ for the ECL barrel and $100 (150)\unit{MeV}$ for the forward (backward) endcap.
Neutral pion candidates with at least one photon being used in the tag-side full reconstruction are discarded.

Events are required to have exactly two oppositely charged particles within the allowed impact parameter range, with one additional track allowed outside the range.
At least one charged pion is required.
If the event contains two charged pions and neutral pion candidates, we search for $\rho^\pm$ candidates.
The charged pion with the lower momentum in the c.m. frame is combined with each neutral pion candidate and a mass vertex fit is performed.
A pair that can be successfully fitted with $\chi^2 < 20$ is accepted as a charged $\rho^\pm$ meson if its invariant mass is between $625$ and $925\cunit{MeV}{/c^2}$.
If multiple candidates are found, the candidate with a mass closest to the nominal $\rho^+$ mass~\cite{PDG-2014} is selected.
Due to the broad $\rho^\pm$ mass range, not all $\tau^- \to \rho^- \nu_\tau$ events are correctly reconstructed.
These events contain two oppositely charged pions and a neutral pion in the final state and are miscategorized in the $\tau^- \to \pi^- \nu_\tau$ channel.
Each event is reconstructed in one of the four reconstruction channels.
In many $\tau^- \to \mu^- \bar{\nu}_\mu \nu_\tau$ events, the momentum of the muon is too low to reach the KLM and thus is not identified as a muon.
In most of these cases, the muon is identified as a pion so that the event is placed in the $\tau^- \to \pi^- \nu_\tau$ sample.

Since $K_L$ mesons do not completely deposit their energy in the detector, charmed $B$ decays with subsequent decays $D \to K_L \pi$ or $D \to K_L \ell \nu_{\ell}$ have the signal's missing-momentum signature.
A $K_L$ candidate is identified as a cluster in the KLM without an associated charged track.
An ECL cluster without an associated charged track is associated with the $K_L$ cluster in the KLM if it lies along the flight path extrapolated from the interaction point to the KLM cluster.
As described below, the extra energy in the ECL is used to determine the signal yield.
Therefore, only events with a $K_L$ without energy deposition in the ECL are vetoed.

\subsection{Extra energy}
We extract the signal yield from a fit to the distribution of the energy deposited in the ECL (\eecl{}) by particles not used in the full reconstruction or by the two remaining charged signal tracks.
To reduce background, the aforementioned photon energy requirements are applied.
For signal decays, there is no additional energy deposition, so \eecl{} peaks strongly at zero.
Misreconstructions of \btag{} lead to a small tail towards higher energy depositions for true signal events.
In contrast, most background decays exhibit non-vanishing extra energy due to the presence of additional neutral particles.

\subsection{Boosted decision trees}
\label{sec:evtsel:bdt}
Final event selection uses requirements on three variables: \lncntnbout{}; missing mass squared ($\mmiss^{2} = \pmiss^{2}$); and the output of the boosted decision tree (BDT).
For each $\tau$ reconstruction mode, one BDT is trained using the TMVA framework~\cite{tmva}.
All use different input variables, background training samples, and BDT growth parameters.
The signal training sample consists of $3 \times 10^{7}$ events out of the complete signal MC sample.
To improve the training, events are required to have $\eecl < 1\unit{GeV}$ and \btag{} is required to have a quality of $\lncntnbout > -7$.
One additional track outside the impact parameter requirement is allowed.

Another $3 \times 10^{7}$ signal events are used for performance tests of the BDT for receiver-operation characteristics (ROC) calculation and overtraining evaluation.

The input variables of the BDT used in the $\tau^- \to e^- \bar{\nu}_e \nu_\tau$ selection are the magnitude of the three-momenta of the pion and electron, the squared lepton-pair momentum transfer $q^2$, $\mmiss^2$, and different combinations of all available four-momenta.
The momentum transfer can be calculated using the fact that both $B$ mesons are at rest in the c.m. frame, which implies $p_{\bsig} = - p_{\btag}$ and $q = p_{\bsig} - p_\pi$.
Due to the low efficiency of the full reconstruction, an additional signal sample is used with $2 \times 10^{7}$ $\tau^- \to e^- \bar{\nu}_e \nu_\tau$ events on the signal side.
The background training sample consists of charmed $B^0$ decays and $B^0 \to X_u \ell \nu_\ell$ decays. 
The input variables are linearly decorrelated before their use in the training.

The background sample used in the $\tau^- \to \pi^- \nu_\tau$ selection BDT contains $b \to c$ decays and semileptonic $b \to u$ decays of $B^0$ mesons. 
Principal-component analysis (PCA)~\cite{tmva} is applied to the input variables.
PCA is an orthogonal transformation which rotates a sample of data points such that the variability along the new axis is maximized.
In this way, the variables are decorrelated.
The input variables are $\mmiss^2$, the missing energy, $q^2$, the absolute three-momentum of \bvis{} in the c.m. frame, combinations of available four-momenta, and the number of unused neutral pions in the event.

The BDT training for the $\tau^- \to \rho^- \nu_\tau$ selection uses the same sample size of $b \to c$ decays, but not semileptonic $b \to u$ decays.
The correlation of the input variables is again reduced by a PCA transformation.
The variables used in the training are $\mmiss^2$, the missing energy, $q^2$, and combinations of the available four-momenta in the decay.

\begin{figure}[htb]
  \includegraphics[width=0.45\textwidth]{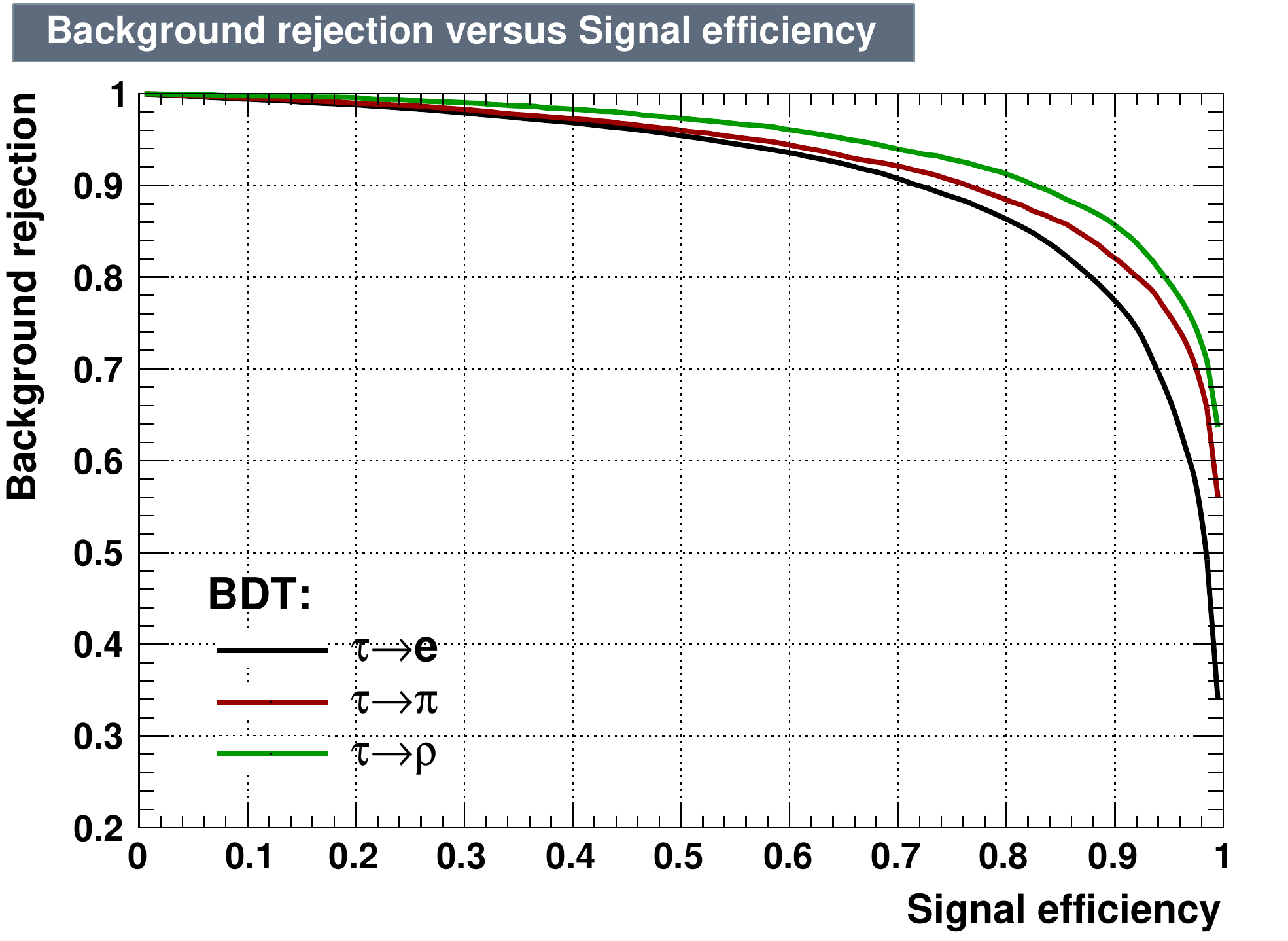}
\caption{Background rejection versus signal efficiency determined on the testing sample, from Belle simulation.}
\label{fig:roc}
\end{figure}

The performance of the three final BDTs is shown in form of the ROC curves in Fig.~\ref{fig:roc}.

\subsection{Final selection}
\label{sec:evtsel:final}
The final selection criteria are determined from MC samples by maximizing individually the expected significance of each single $\tau$ reconstruction mode.
We perform a scan over three variables simultaneously to obtain the optimal selection: \lncntnbout, $\mmiss^{2}$, and the BDT output.
We require $\lncntnbout > -7$ for the leptonic $\tau$ reconstruction and $\lncntnbout > -5$ for the hadronic $\tau$ reconstruction, as shown in Fig.~\ref{fig:tagside}.
A minimum requirement on the $\mmiss^{2}$ is applied to reject semileptonic $B \to \pi \ell \nu$ events, which have the same final state as the signal decay: since no energy is deposited in the ECL, decays of this type peak at zero extra energy.
Also, as there is only a single neutrino in these decays, $\mmiss^2$ peaks at zero, unlike the case for signal decays, which contain at least two neutrinos.
We require $\mmiss^{2}$ to be greater than $2.2\cunit{GeV}{/c^4}$ in the electron channel, $0.0\cunit{GeV}{/c^4}$ in the pion channel, and $0.6\cunit{GeV}{/c^2}$ in the $\rho$ channel.
The BDT output is the last variable used in the scan.
The expected significance is calculated as $\sqrt{ -2 \ln (\mathcal{L}_0 / \mathcal{L}_1) }$; the likelihood is given by $\mathcal{L}_k = \prod^{i}_{\rm bins} \mathrm{Poisson}( n_{\rm obs} | n_{\rm bg} + k \cdot n_{\rm sig} )$, where $n_{\rm obs} = n_{\rm bg} + n_{\rm sig}$ is the best estimate from the MC samples.

The efficiency of the final selection is determined from MC to be $4.57 \times 10^{-4}$.
The dominant reconstructed $\tau$ decay modes and their relative occurrences are listed in Table~\ref{tab:signalcomp}.

\begin{table}[htb]
  \caption{Signal reconstruction by $\tau$ decay modes. Percentages are obtained from signal MC and sum to $100\%$.}
\label{tab:signalcomp}
\begin{tabular}
 {@{\hspace{0.5cm}}l@{\hspace{0.5cm}}  @{\hspace{0.5cm}}c@{\hspace{0.5cm}}}
\hline \hline
$\tau^-$ decay & Relative Occurrence (\%) \\
\hline
$\rho^- \nu_\tau$ & $29.54$ \\
$e^- \bar{\nu}_e \nu_\tau$ & $29.43$ \\
$\pi^- \nu_\tau$ & $16.70$ \\
$\mu^- \bar{\nu}_\mu \nu_\tau$ & $13.21$ \\
$a_1^- \nu_\tau$ & $8.72$ \\
other & $2.4$ \\
\hline \hline
\end{tabular}
\end{table}

The dominant background in the low-\eecl{} region arises from $B^0 \to D^{(*)} \ell \nu_\ell$ and $B^0 \to D^{(*)} \rho$ decays with a subsequent decay of $D \to K_L \pi$.
The $K_L$ is undetected in these cases and the resulting decay signature resembles that of the signal.
No explicit selection is available to further suppress decays of this type.

\section{Systematic uncertainties}
In the computation of the significance level and upper limit, systematic uncertainties are included in the likelihood as nuisance parameters.
The likelihood is built from probability density functions (PDFs) determined from MC predictions of each background sample, as described in Section~\ref{sec:result}.
All systematic uncertainties are assumed to be Gaussian-distributed and are evaluated at one standard deviation ($\sigma$).

Uncertainties of the particle identification and of the correction factor needed for the full reconstruction efficiency are included as a flat effect over all bins in \eecl{}.
All other uncertainties are included in a bin-by-bin fashion.
A constant uncertainty of $0.35\%$ has been determined for each charged track with $p_t > 0.2\cunit{GeV}{/c}$.
Tracks below that threshold have to be treated differently depending on the track momentum~\cite{Dungel_slowpi}.
The uncertainty on the number of produced $B$-meson pairs is $1.4\%$.
The uncertainty due to the $K_L$ veto is determined by varying the $K_L$ efficiency by its uncertainty.
The branching fractions of the dominant backgrounds are varied by their errors stated in Ref.~\cite{PDG-2014} to determine the effect on the MC prediction.
The uncertainty on the correction factor of the tagside reconstruction is determined in Ref.~\cite{alexei_bulnu} and applied to the samples.
The discrepancy between inclusive and exclusive $|V_{ub}|$ measurements has been included as a flat but asymmetric uncertainty in the $B \to X_u \ell \nu$ sample of $(^{+5}_{-15})\%$.
An uncertainty of $\pm 10\%$ is applied to the branching fractions in the MC sample of rare $b \to s$ and other rare $B$ decays.
Additionally, decays of type $B \to X_u \tau \nu$ are present in the final event selection.
The contribution to the \eecl{} distribution is evaluated from the MC sample assuming a $\mathcal{B}( B^0 \to \rho^+ \tau^- \nu_\tau ) = 1.5 \times 10^{-4}$ and found to be small; a relative uncertainty of $\pm 50\%$ is applied.
Statistical uncertainties in the PDF shape due to finite MC sample size are included in a way similar to the approach by Barlow and Beeston~\cite{barlow-beeston}.
Instead of using one Poisson constraint per background sample per bin per $\tau$ decay channel, only one constraint term per bin per channel is used.
The uncertainty introduced by this approximation is negligible for bins with non-vanishing content and reduces the amount of computation time needed.
Instead of the finite MC uncertainty, the fit error is included as a systematic uncertainty for the dominant $b \to c$ contribution.
The theoretical uncertainties of the signal form factors $f^+$ and $f^0$ are included by generating additional signal MC with one form factor fixed and the other varied by its $1\sigma$ uncertainty.
The relative uncertainties determined in this way are combined into a single uncertainty estimate.
The systematic uncertainties due to the tracking efficiency and particle identification affect only the overall efficiency and are only included in the calculation of the upper limit.
The relative effect on the branching fraction is determined by repeatedly fitting modified PDFs to data.
The PDFs are modified by replacing each background contribution with the respective contribution where the systematic effect is applied.
For each systematic uncertainty, two fits are performed for the positive and negative deviation.
The maximum, absolute deviation is quoted in Table~\ref{tab:syseffects}.

\begin{table}[htb]
\caption{Effects of the systematic uncertainties on the branching fraction.}
\label{tab:syseffects}
\begin{tabular}
 {@{\hspace{0.5cm}}l@{\hspace{0.5cm}}  @{\hspace{0.5cm}}c@{\hspace{0.5cm}}}
\hline \hline
Source & Relative error (\%) \\
\hline
	Particle ID & $2.4$ \\
	Track efficiency & $0.7$ \\
	N(\BB) & $1.4$ \\
	$K_L$ veto & $3.2$ \\
	BG $\mathcal{B}$ & $2.8$ \\
	Tagside & $4.6$ \\
	$|V_{ub}|$ & $2.8$ \\
	Rare processes & $2.0$ \\
	$B \to X_u \tau \nu$ & $2.2$ \\
	Background fit & $0.2$ \\
	Signal model & $1.8$ \\
\hline
	Total & $8.3$ \\
\hline \hline
\end{tabular}
\end{table}

\section{Result}
\label{sec:result}
\begin{figure*}[htb]
  \begin{tabular}{ccc}
	\begin{subfigure}[t]{0.45\textwidth}
	  \includegraphics[width=1.\textwidth]{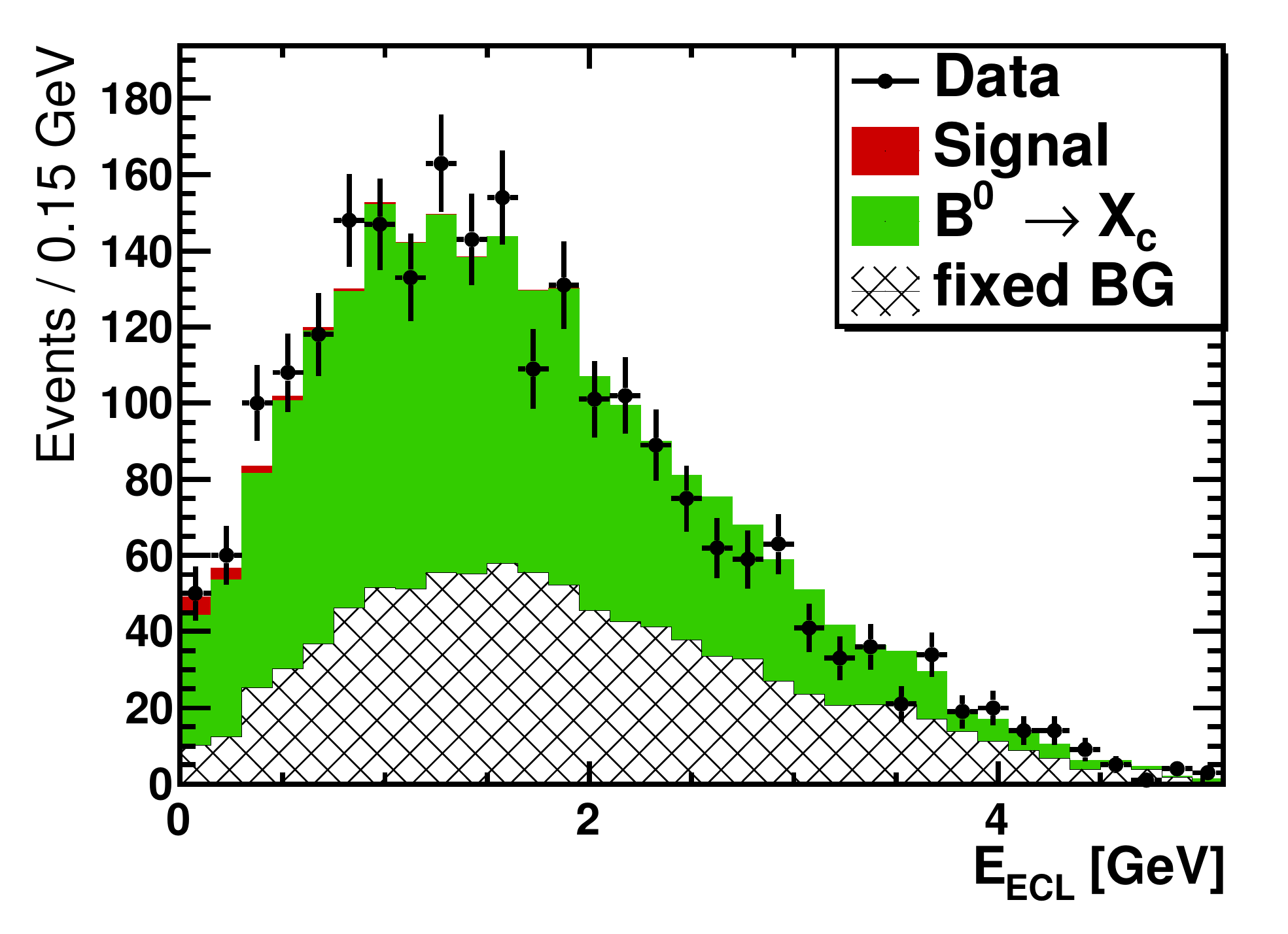}
	  \caption{$\tau \to e \nu \nu$}
	\end{subfigure}
	&
	\begin{subfigure}[t]{0.45\textwidth}
	  \includegraphics[width=1.\textwidth]{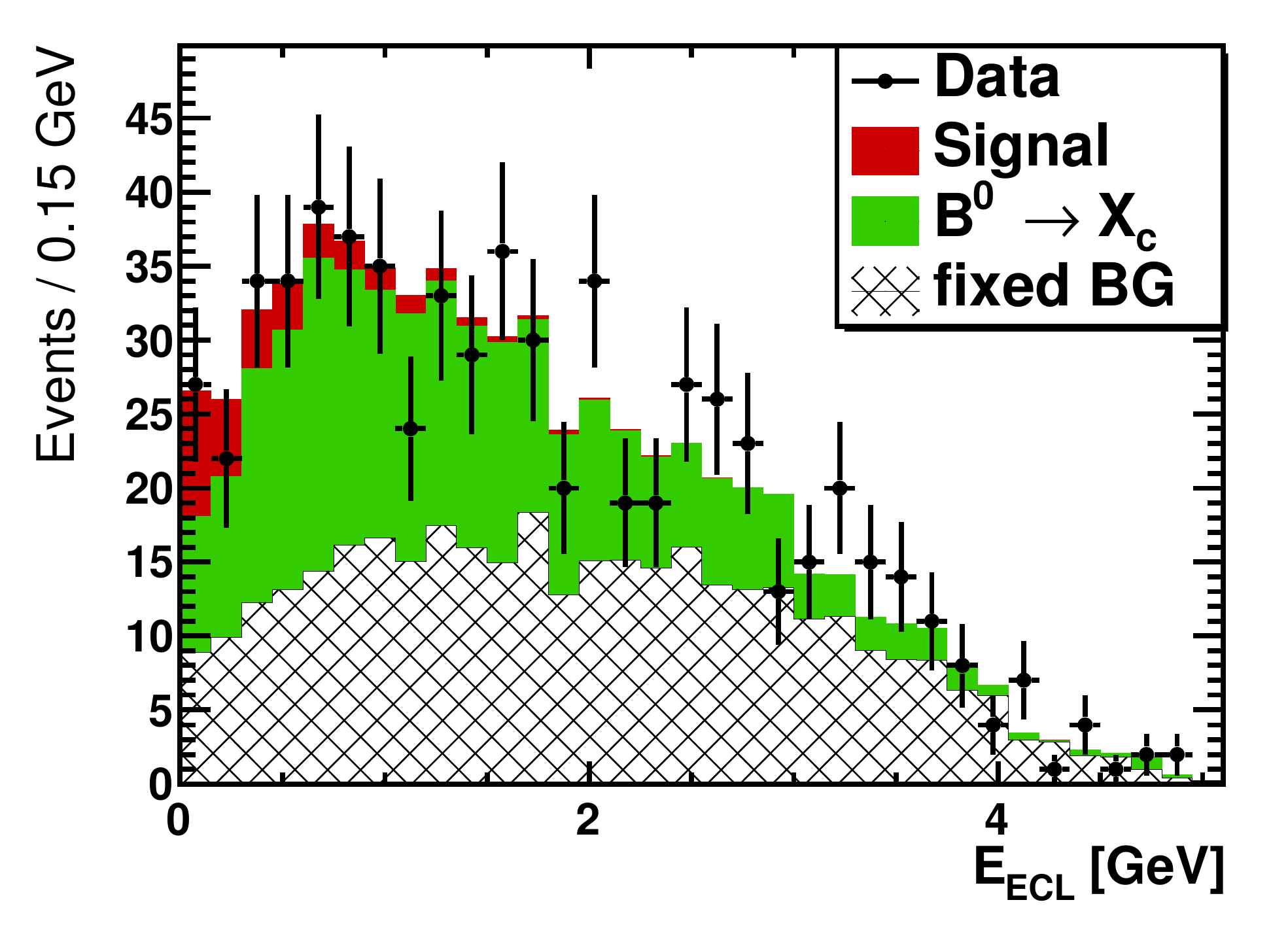}
	  \caption{$\tau \to \pi \nu$}
	\end{subfigure}
	\\
	\multicolumn{2}{c}{
	\begin{subfigure}[t]{0.45\textwidth}
	  \includegraphics[width=1.\textwidth]{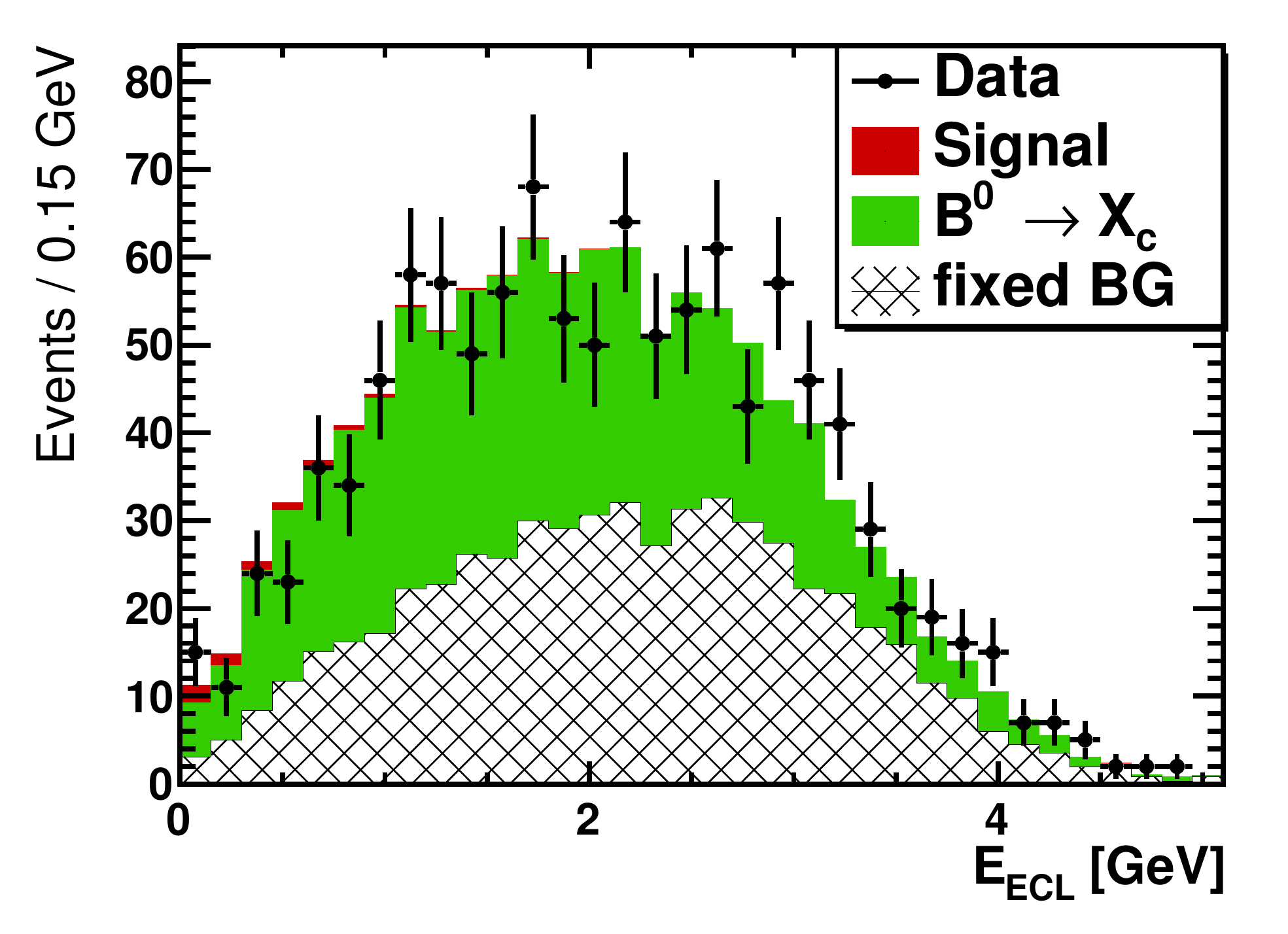}
	  \caption{$\tau \to \rho \nu$}
	\end{subfigure}
  }
  \end{tabular}
\caption{Distributions of \eecl{} in the three $\tau$ reconstruction modes. The signal and $b \to c$ contributions are scaled according to the fit result.}
\label{fig:fit}
\end{figure*}

A binned maximum likelihood fit is performed to \eecl{} in bins of $0.15\unit{GeV}$.
Due to similar shapes in the background predictions, all background contributions except for the dominant $b \to c$ transitions are fixed to the MC prediction.
Possible errors introduced by this approach are accounted for as systematic uncertainties.
The fit is performed simultaneously in all three reconstruction modes.
The signal strength parameter $\mu$ is constrained between the three modes while the background contributions of the three reconstruction modes are floating parameters.
The fit result of the $B^0 \to X_c$ background contribution agrees well with the prediction obtained from the MC sample.
The signal strength has been chosen such that $\mu = 1.0$ corresponds to $\mathcal{B}( \pitaunu ) = 1.0 \times 10^{-4}$.
We obtain a best fit of $\mu = 1.52 \pm 0.72$, corresponding to $51.9 \pm 24.3$ signal events.
The fit results by $\tau$ reconstruction mode are listed in Table~\ref{tab:signalyield}.
\begin{table}[htb]
\caption{Fit results for signal yield. Total and split by $\tau$ reconstruction mode.}
\label{tab:signalyield}
\begin{tabular}
  {@{\hspace{0.5cm}}c@{\hspace{0.5cm}}  @{\hspace{0.5cm}}r@{\hspace{0.5cm}}}
  \hline \hline
  Mode & Signal Yield \\
  \hline
  $e$ & $13.2 \pm 6.2$ \\
  $\pi$ & $30.6 \pm 14.3$ \\
  $\rho$ & $8.1 \pm 3.8$ \\
  \hline
  Total & $51.9 \pm 24.3$\\
  \hline \hline
\end{tabular}

\end{table}
The \eecl{} distribution and fit results are shown in Fig.~\ref{fig:fit}.

The significance of the measurement is obtained from a pseudo MC study.
A test statistic based on the profile likelihood ratio is used.
The likelihood is built in bins of $0.15\unit{GeV}$ in \eecl{}.
The binned likelihood is given by
\begin{equation}
  \mathcal{L} = \prod\limits_c~\prod\limits_b \mathrm{Poisson}( n_{cb} | \nu_{cb} ) \cdot \prod\limits_{p \in \mathbb{P}} f_p( a_p | \alpha_p ),
  \label{eq:likelihood}
\end{equation}
where the indices $c$ and $b$ label the reconstruction channel and bin in \eecl, respectively, and
$\mathbb{P}$ denotes the set of systematic uncertainties $p$ that are included as nuisance parameters $\alpha_p$ in the calculation of the number of expected events $\nu_{cb}$ per channel per bin.
The nuisance parameters are parametrized as a relative effect on the nominal template prediction, assumed to be Gaussian-distributed with the nominal value being the global observable $a_p$.
The number of events in the background-only hypothesis is determined from MC simulation and a fit to data for the dominant $b \to c$ background.
The likelihood is constructed using the \texttt{HistFactory} tool in the RooStats package~\cite{roostats,histfactory}.

The distribution of the test statistic is obtained by pseudo-experiments.
A full frequentist approach is used in both the computations of the significance level and the upper limit.
First, the likelihood is fitted to data to obtain the maximum likelihood estimates (MLEs) of all nuisance parameters on data.
In each pseudo-experiment generation, the nuisance parameters are fixed to their respective MLE.
In the subsequent maximization of the likelihood, the nuisance parameters are free parameters.
The global observables are randomized in each pseudo-experiment.

Using pseudo-experiments, the \pvalue{} of the background-only hypothesis for data is determined and the significance level $Z$ is computed in terms of standard deviations as
\[
  Z = \Phi^{-1} \left( 1 - p \right),
\]
where $\Phi^{-1}$ is the cumulative distribution function of the standard normal Gaussian.

We observe a signal significance of $2.8\sigma$, not including systematic uncertainties in the calculation.
Including all relevant systematic effects results in a significance of $2.4\sigma$.
For this result, the test statistic has been computed on $10\,000$ background-only pseudo-experiments.

Given the level of significance of these results, we invert the hypothesis test and compute an upper limit on the branching fraction.
pseudo-experiments are generated for different signal strength parameters for both signal-plus-background and background-only hypotheses in order to obtain $CL_{s+b}$ and $CL_{b}$, respectively.
The upper limit is then computed using $CL_s = CL_{s+b} / CL_b$~\cite{cls_Read}, where a scan over reasonable signal strength parameter values is performed.
At each step, $10\,000$ pseudo-experiments have been evaluated for both hypotheses.

At the 90\% confidence level, we obtain an upper limit of \mbox{$\br{\pitaunu} < 2.5 \times 10^{-4}$}.
The upper limit at the 95\% confidence level has been computed to \mbox{$\br{\pitaunu} < 2.8 \times 10^{-4}$}.
This result is the first result on $\br{\pitaunu}$ and is in good agreement with the SM prediction.


\begin{acknowledgments}
We thank the KEKB group for the excellent operation of the
accelerator; the KEK cryogenics group for the efficient
operation of the solenoid; and the KEK computer group,
the National Institute of Informatics, and the
PNNL/EMSL computing group for valuable computing
and SINET4 network support.  We acknowledge support from
the Ministry of Education, Culture, Sports, Science, and
Technology (MEXT) of Japan, the Japan Society for the
Promotion of Science (JSPS), and the Tau-Lepton Physics
Research Center of Nagoya University;
the Australian Research Council;
Austrian Science Fund under Grant No.~P 22742-N16 and P 26794-N20;
the National Natural Science Foundation of China under Contracts
No.~10575109, No.~10775142, No.~10875115, No.~11175187, and  No.~11475187;
the Chinese Academy of Science Center for Excellence in Particle Physics;
the Ministry of Education, Youth and Sports of the Czech
Republic under Contract No.~LG14034;
the Carl Zeiss Foundation, the Deutsche Forschungsgemeinschaft
and the VolkswagenStiftung;
the Department of Science and Technology of India;
the Istituto Nazionale di Fisica Nucleare of Italy;
the WCU program of the Ministry of Education, National Research Foundation (NRF)
of Korea Grants No.~2011-0029457,  No.~2012-0008143,
No.~2012R1A1A2008330, No.~2013R1A1A3007772, No.~2014R1A2A2A01005286,
No.~2014R1A2A2A01002734, No.~2015R1A2A2A01003280 , No. 2015H1A2A1033649;
the Basic Research Lab program under NRF Grant No.~KRF-2011-0020333,
Center for Korean J-PARC Users, No.~NRF-2013K1A3A7A06056592;
the Brain Korea 21-Plus program and Radiation Science Research Institute;
the Polish Ministry of Science and Higher Education and
the National Science Center;
the Ministry of Education and Science of the Russian Federation and
the Russian Foundation for Basic Research;
the Slovenian Research Agency;
the Basque Foundation for Science (IKERBASQUE) and
the Euskal Herriko Unibertsitatea (UPV/EHU) under program UFI 11/55 (Spain);
the Swiss National Science Foundation; the National Science Council
and the Ministry of Education of Taiwan; and the U.S.\
Department of Energy and the National Science Foundation.
This work is supported by a Grant-in-Aid from MEXT for
Science Research in a Priority Area (``New Development of
Flavor Physics'') and from JSPS for Creative Scientific
Research (``Evolution of Tau-lepton Physics'').
\end{acknowledgments}

\bibliography{pitaunu}

\end{document}

%% file: author.tex
\noaffiliation
\affiliation{University of the Basque Country UPV/EHU, 48080 Bilbao}
\affiliation{University of Bonn, 53115 Bonn}
\affiliation{Budker Institute of Nuclear Physics SB RAS, Novosibirsk 630090}
\affiliation{Faculty of Mathematics and Physics, Charles University, 121 16 Prague}
\affiliation{Chonnam National University, Kwangju 660-701}
\affiliation{University of Cincinnati, Cincinnati, Ohio 45221}
\affiliation{Deutsches Elektronen--Synchrotron, 22607 Hamburg}
\affiliation{Justus-Liebig-Universit\"at Gie\ss{}en, 35392 Gie\ss{}en}
\affiliation{II. Physikalisches Institut, Georg-August-Universit\"at G\"ottingen, 37073 G\"ottingen}
\affiliation{SOKENDAI (The Graduate University for Advanced Studies), Hayama 240-0193}
\affiliation{Hanyang University, Seoul 133-791}
\affiliation{University of Hawaii, Honolulu, Hawaii 96822}
\affiliation{High Energy Accelerator Research Organization (KEK), Tsukuba 305-0801}
\affiliation{IKERBASQUE, Basque Foundation for Science, 48013 Bilbao}
\affiliation{Indian Institute of Technology Bhubaneswar, Satya Nagar 751007}
\affiliation{Indian Institute of Technology Guwahati, Assam 781039}
\affiliation{Indian Institute of Technology Madras, Chennai 600036}
\affiliation{Indiana University, Bloomington, Indiana 47408}
\affiliation{Institute of High Energy Physics, Chinese Academy of Sciences, Beijing 100049}
\affiliation{Institute of High Energy Physics, Vienna 1050}
\affiliation{Institute for High Energy Physics, Protvino 142281}
\affiliation{INFN - Sezione di Torino, 10125 Torino}
\affiliation{J. Stefan Institute, 1000 Ljubljana}
\affiliation{Kanagawa University, Yokohama 221-8686}
\affiliation{Institut f\"ur Experimentelle Kernphysik, Karlsruher Institut f\"ur Technologie, 76131 Karlsruhe}
\affiliation{Kennesaw State University, Kennesaw GA 30144}
\affiliation{King Abdulaziz City for Science and Technology, Riyadh 11442}
\affiliation{Department of Physics, Faculty of Science, King Abdulaziz University, Jeddah 21589}
\affiliation{Korea Institute of Science and Technology Information, Daejeon 305-806}
\affiliation{Korea University, Seoul 136-713}
\affiliation{Kyungpook National University, Daegu 702-701}
\affiliation{\'Ecole Polytechnique F\'ed\'erale de Lausanne (EPFL), Lausanne 1015}
\affiliation{Faculty of Mathematics and Physics, University of Ljubljana, 1000 Ljubljana}
\affiliation{Ludwig Maximilians University, 80539 Munich}
\affiliation{Luther College, Decorah, Iowa 52101}
\affiliation{University of Maribor, 2000 Maribor}
\affiliation{Max-Planck-Institut f\"ur Physik, 80805 M\"unchen}
\affiliation{School of Physics, University of Melbourne, Victoria 3010}
\affiliation{Moscow Physical Engineering Institute, Moscow 115409}
\affiliation{Moscow Institute of Physics and Technology, Moscow Region 141700}
\affiliation{Graduate School of Science, Nagoya University, Nagoya 464-8602}
\affiliation{Kobayashi-Maskawa Institute, Nagoya University, Nagoya 464-8602}
\affiliation{Nara Women's University, Nara 630-8506}
\affiliation{National Central University, Chung-li 32054}
\affiliation{National United University, Miao Li 36003}
\affiliation{Department of Physics, National Taiwan University, Taipei 10617}
\affiliation{H. Niewodniczanski Institute of Nuclear Physics, Krakow 31-342}
\affiliation{Niigata University, Niigata 950-2181}
\affiliation{Novosibirsk State University, Novosibirsk 630090}
\affiliation{Osaka City University, Osaka 558-8585}
\affiliation{Pacific Northwest National Laboratory, Richland, Washington 99352}
\affiliation{University of Pittsburgh, Pittsburgh, Pennsylvania 15260}
\affiliation{University of Science and Technology of China, Hefei 230026}
\affiliation{Soongsil University, Seoul 156-743}
\affiliation{Sungkyunkwan University, Suwon 440-746}
\affiliation{School of Physics, University of Sydney, NSW 2006}
\affiliation{Department of Physics, Faculty of Science, University of Tabuk, Tabuk 71451}
\affiliation{Tata Institute of Fundamental Research, Mumbai 400005}
\affiliation{Excellence Cluster Universe, Technische Universit\"at M\"unchen, 85748 Garching}
\affiliation{Department of Physics, Technische Universit\"at M\"unchen, 85748 Garching}
\affiliation{Toho University, Funabashi 274-8510}
\affiliation{Tohoku University, Sendai 980-8578}
\affiliation{Earthquake Research Institute, University of Tokyo, Tokyo 113-0032}
\affiliation{Department of Physics, University of Tokyo, Tokyo 113-0033}
\affiliation{Tokyo Institute of Technology, Tokyo 152-8550}
\affiliation{Tokyo Metropolitan University, Tokyo 192-0397}
\affiliation{Utkal University, Bhubaneswar 751004}
\affiliation{CNP, Virginia Polytechnic Institute and State University, Blacksburg, Virginia 24061}
\affiliation{Wayne State University, Detroit, Michigan 48202}
\affiliation{Yamagata University, Yamagata 990-8560}
\affiliation{Yonsei University, Seoul 120-749}
  \author{P.~Hamer}\affiliation{II. Physikalisches Institut, Georg-August-Universit\"at G\"ottingen, 37073 G\"ottingen} 
  \author{A.~Frey}\affiliation{II. Physikalisches Institut, Georg-August-Universit\"at G\"ottingen, 37073 G\"ottingen} 
  \author{A.~Abdesselam}\affiliation{Department of Physics, Faculty of Science, University of Tabuk, Tabuk 71451} 
  \author{I.~Adachi}\affiliation{High Energy Accelerator Research Organization (KEK), Tsukuba 305-0801}\affiliation{SOKENDAI (The Graduate University for Advanced Studies), Hayama 240-0193} 
  \author{H.~Aihara}\affiliation{Department of Physics, University of Tokyo, Tokyo 113-0033} 
  \author{S.~Al~Said}\affiliation{Department of Physics, Faculty of Science, University of Tabuk, Tabuk 71451}\affiliation{Department of Physics, Faculty of Science, King Abdulaziz University, Jeddah 21589} 
  \author{K.~Arinstein}\affiliation{Budker Institute of Nuclear Physics SB RAS, Novosibirsk 630090}\affiliation{Novosibirsk State University, Novosibirsk 630090} 
  \author{D.~M.~Asner}\affiliation{Pacific Northwest National Laboratory, Richland, Washington 99352} 
  \author{T.~Aushev}\affiliation{Moscow Institute of Physics and Technology, Moscow Region 141700} 
  \author{R.~Ayad}\affiliation{Department of Physics, Faculty of Science, University of Tabuk, Tabuk 71451} 
% ...
\collaboration{The Belle Collaboration}